\documentclass[twocolumn,showpacs,amsfonts,aps,prc,floatfix,superscriptaddress
]{revtex4-1}

\usepackage[colorlinks=true, pdfstartview=FitV, linkcolor=red, citecolor=blue, urlcolor=blue]{hyperref}
\usepackage{amsmath}
\usepackage{bm}
\usepackage{graphicx}

\begin{document}

\title{
Off-equilibrium corrections to energy and conserved charge densities in the relativistic fluid in heavy-ion collisions
}

\author{Akihiko Monnai}
\email[]{akihiko.monnai@ipht.fr}
\affiliation{Institut de Physique Th\'{e}orique, CNRS, CEA/Saclay, F-91191
Gif-sur-Yvette, France}
\affiliation{KEK Theory Center, Institute of Particle and Nuclear Studies, \\
High Energy Accelerator Research Organization (KEK),
1-1, Ooho, Tsukuba, Ibaraki 305-0801, Japan}
\date{\today}

\begin{abstract}
Dissipative processes in relativistic fluids are known to be important in the analyses of the hot QCD matter created in high-energy heavy-ion collisions. In this work, I consider dissipative corrections to energy and conserved charge densities, which are conventionally assumed to be vanishing but could be finite. Causal dissipative hydrodynamics is formulated in the presence of those dissipative currents. The relation between hydrodynamic stability and transport coefficients is discussed. I then study their phenomenological consequences on the observables of heavy-ion collisions in numerical simulations. It is shown that particle spectra and elliptic flow can be visibly modified.
\end{abstract}

\pacs{25.75.-q, 25.75.Nq, 25.75.Ld}

\maketitle

\section{Introduction}
\label{sec1}
\vspace*{-2mm}

Two decades of heavy-ion programs at Relativistic Heavy Ion Collider (RHIC) at Brookhaven National Laboratory \cite{Adcox:2004mh,Adams:2005dq,Back:2004je,Arsene:2004fa} have established that relativistic hydrodynamics is a powerful framework to analyze the collective properties of the quark-gluon plasma (QGP), a QCD matter in the deconfined phase \cite{Yagi:2005yb}. The fact is also confirmed at higher energies at Large Hadron Collider (LHC) in European Organization for Nuclear Research \cite{Aamodt:2010pa,ATLAS:2011ah,Chatrchyan:2012wg}. Experimental observation of large hadronic flow harmonics, defined with the Fourier coefficients of azimuthal momentum distribution \cite{Ollitrault:1992bk,Poskanzer:1998yz}, are considered to be an evidence for the existence of the nearly-perfect QGP fluid. The precision analyses in the past ten years have revealed that they are much better described with the help of viscosity \cite{MulRug98b}, which takes account of the deviation of the system from local thermal equilibrium \cite{Wang:2016opj}. Sophisticated versions of the relativistic dissipative hydrodynamic models have been analytically and numerically solved and used as a quantitative and dynamical description of the hot and dense QCD matter, for which first principle calculations may have difficulties.

Of different types of dissipative currents, shear stress tensor $\pi^{\mu \nu}$, which is the response to deformation of the target system, is the first to be introduced to the heavy-ion phenomenology \cite{Romatschke:2007mq,Chaudhuri:2007zm,Luzum:2008cw,Song:2007fn} because it is non-vanishing in the conformal limit and a fairly large contribution is expected in heavy-ion systems where its shape is deformed very rapidly. Bulk pressure $\Pi$, which is the response to expansion and compression, is the next to be recognized \cite{Denicol:2009am,Song:2009rh,Bozek:2009dw}, as even though it is vanishing in the conformal limit it can be as large as shear viscosity in the vicinity of quark-hadron crossover owing to the broken scale invariance \cite{Kharzeev:2007wb}. This is expected to have a visible effect at freeze-out, where hydrodynamic flow is converted into particles \cite{Monnai:2009ad, Denicol:2009am}. It has been established that those viscous corrections are essential for the quantitative understanding of experimental data \cite{Ryu:2015vwa}. Baryon diffusion $V_B^\mu$, which is the response to the gradient in the fugacity, is beginning to attract attention \cite{Monnai:2012jc,Shen:2017ruz,Denicol:2018wdp} because baryon-rich matter is expected to be produced in the Beam Energy Scan program at RHIC.

The tensor decompositions of the off-equilibrium energy-momentum tensor $\delta T^{\mu \nu}$ and conserved charge current $\delta N_J^\mu$ leads to dissipative currents $w = \delta T^{\mu \nu} u_\mu u_\nu$ and $v_J = \delta N_J^\mu u_\mu$, which can be interpreted at the linear order as the responses to changes of inverse temperature and fugacity of the $J$-th conserved charge, respectively. They are conventionally set to be vanishing using the so-called Landau matching conditions \cite{Israel:1979wp}, which are introduced to make use of the equation of state for closing the hydrodynamic set of equations. However, this might not be a strong motivation to neglect those dissipative quantities because their constitutive equations can be derived and the system can still be uniquely solved by calculating equilibrium energy density for the equation of state. The matching conditions are also argued from the viewpoint of thermodynamic stability \cite{Monnai:2010qp}, but it is not clear if non-vanishing first-order derivatives of the entropy density with respect to dissipative currents imply instability because the dissipative currents are not macroscopic variables that can be determined from the fields of flow, temperature and chemical potential at given times. 
The microscopic origin of the Landau matching condition has not been sufficiently discussed so far except for the one based on the renormalization group technique \cite{Tsumura:2006hn,Tsumura:2009vm,Tsumura:2012ss}. A generalized version of the matching condition and finite corrections to the energy and number densities are discussed in Ref.~\cite{Eckart:1940te, Osada:2009cc, Osada:2011gx, Osada:2014bza}. The dissipative correction to the energy density is also argued from the viewpoint of causality at non-linear order when the fluid is coupled to gravity \cite{Bemfica:2017wps}. It would thus be important to investigate theoretical and phenomenological consequences of those dissipative quantities and to see if they would cause instability by keeping those variables finite and performing extended analytical and numerical analyses. It is note-worthy that conservation laws can still be imposed to the system and are not affected by the presence of those dissipative currents. 

In this paper, I study the effects of off-equilibrium corrections to energy and charge densities in a relativistic fluid analytically and numerically.
The paper is organized as follows. In Sec.~\ref{sec:rdh}, I discuss relativistic dissipative hydrodynamics with off-equilibrium energy and charge densities. The second-order constitutive equations are derived and then investigated theoretically. Sec.~\ref{sec:num} is devoted to the numerical estimations of the effects of those dissipative currents in a system of ultrarelativistic heavy-ion collision. Discussion and conclusions are presented in Sec.~\ref{sec:disconc}. The natural unit $c = \hbar = k_B = 1$ and the mostly-minus Minkowski metric $g^{\mu \nu} = \mathrm{diag}(+,-,-,-)$ are used throughout this paper.

\section{Relativistic dissipative hydrodynamics}
\label{sec:rdh}
\vspace*{-2mm}

I discuss theoretical formulation of causal and stable relativistic hydrodynamics in the presence of dissipative processes including the off-equilibrium corrections to energy and net charge densities. The Israel-Stewart-type second order theory is employed for the derivation of the dissipative hydrodynamic equations of motion for isotropic media \cite{Israel:1979wp, Monnai:2010qp}.

\subsection{Equations of Motion}

Energy-momentum tensor and net charge currents are the conserved quantities in a standard hydrodynamic system. In the absence of dissipative processes, they are described by the energy density $e$, the hydrostatic pressure $P$ and the conserved charge densities $n_J$ $(J = 1, ..., N)$ when tensor decomposition is performed with respect to the flow $u^\mu$. On the other hand, the off-equilibrium corrections $\delta T^{\mu \nu}$ and $\delta N_J^\mu$ introduce $10+4N$ additional variables corresponding to to the number of unknowns in the respective tensors. The decomposition thus reads
\begin{eqnarray}
T^{\mu \nu} &=& (e + w) u^\mu u^\nu - (P+ \Pi) \Delta^{\mu \nu} \nonumber \\
&+& W^\mu u^\nu + W^\nu u^\mu + \pi^{\mu \nu}, \\
N_J^\mu &=& (n_J + v_J) u^\mu + V_J^\mu ,
\end{eqnarray}
where the bulk pressure $\Pi$, the energy dissipation current $W^\mu$, the shear stress tensor $\pi^{\mu \nu}$, and the charge diffusion currents $V_J^\mu$ are the conventional dissipative currents. $w$ and $v_J$ are the off-equilibrium corrections to the energy and the charge densities. $\Delta^{\mu \nu} = g^{\mu \nu} - u^\mu u^\nu$ is the projection operator for the space-like components. Note that the condition $\delta T^\mu_{\ \mu} = 0$, if imposed, still leaves room for $\Pi = w/3 \neq 0$ that is non-vanishing. 

Energy-momentum conservation $\partial_\mu T^{\mu \nu} = 0$ and charge conservations $\partial_\mu N_J^\mu = 0$ provide $4+N$ equations of motion. In addition to the equation of state, one needs constitutive equations to determine the off-equilibrium quantities. They are usually derived from the law of increasing entropy $\partial_\mu s^\mu \geq 0$. Relativistic Navier-Stokes equations are obtained at the lowest order in the expansion of $s^\mu$ in terms of the dissipative quantities. However, they are known to permit superluminal transmission of information and also have unstable modes \cite{Hiscock:1983zz,Hiscock:1985zz}. See Refs.~\cite{Israel:1976tn,Kostadt:2000ty,Geroch:2001xs,Van:2007pw,Van:2011zz,Van:2011yn,Becattini:2014yxa} for further related discussions. In this study, I employ the Israel-Stewart prescription and introduce the second-order terms to the constitutive equations assuming the Grad momentum expansion of the phase-space distribution following Refs.~\cite{Israel:1979wp,Monnai:2010qp}. By keeping $w$ and $v$, the full second-order constitutive relations for the scalar dissipative processes are:
\begin{eqnarray}
w &=& \zeta_{w w} D\frac{1}{T} - \zeta_{w v} D\frac{\mu_J}{T} + \zeta_{w \Pi} \frac{1}{T} \nabla _\mu u^\mu \nonumber \\
&-& \tau_w D w - \tau_{wv_J} D v_J - \tau_{w \Pi} D \Pi  \nonumber \\
&+& \chi_{ww}^{aJ} w D\frac{\mu _J}{T} + \chi_{ww}^b w D \frac{1}{T} + \chi_{ww}^c w \nabla _\mu u^\mu \nonumber \\
&+& \chi_{w \Pi}^{aJ} \Pi D\frac{\mu _J}{T} + \chi_{w \Pi}^b \Pi D \frac{1}{T} + \chi_{w \Pi}^c \Pi \nabla _\mu u^\mu \nonumber \\
&+& \chi_{w v_J}^{aK} v_J D\frac{\mu _K}{T} + \chi_{w v_J}^b v_J D \frac{1}{T} + \chi_{w v_J}^c v_J \nabla _\mu u^\mu \nonumber \\
&+& \chi_{w W}^{aJ} W_\mu \nabla ^\mu \frac{\mu _J}{T} + \chi_{w W}^b W_\mu \nabla ^\mu \frac{1}{T} \nonumber \\ 
&+& \chi_{w W}^c W_\mu D u ^\mu + \chi_{wW}^d \nabla ^\mu W_\mu \nonumber \\
&+& \chi_{w V_J}^{aK} V^J_\mu \nabla ^\mu \frac{\mu_{K}}{T} + \chi_{w V_J}^b V^J_\mu \nabla ^\mu \frac{1}{T} \nonumber \\
&+& \chi_{w V_J}^c V^J_\mu D u ^\mu + \chi_{w V_J}^d \nabla ^\mu V^J_\mu \nonumber \\
&+& \chi_{w\pi} \pi _{\mu \nu} \nabla ^{\langle \mu} u^{\nu \rangle},
\end{eqnarray}
\begin{eqnarray}
\Pi &=& -\zeta_{\Pi \Pi} \frac{1}{T} \nabla _\mu u^\mu - \zeta_{\Pi w} D\frac{1}{T} + \zeta_{\Pi v} D\frac{\mu_J}{T} \nonumber \\
&-& \tau_{\Pi} D \Pi - \tau_{\Pi w} D w - \tau_{\Pi v_J} D v_J \nonumber \\
&+& \chi_{\Pi w}^{aJ} w D\frac{\mu _J}{T} + \chi_{\Pi w}^b w D \frac{1}{T} + \chi_{\Pi w}^c w \nabla _\mu u^\mu \nonumber \\
&+& \chi_{\Pi \Pi}^{aJ} \Pi D\frac{\mu _J}{T} + \chi_{\Pi \Pi}^b \Pi D \frac{1}{T} + \chi_{\Pi \Pi}^c \Pi \nabla _\mu u^\mu \nonumber \\
&+& \chi_{\Pi v_J}^{aK} v_J D\frac{\mu _K}{T} + \chi_{\Pi v_J}^b v_J D \frac{1}{T} + \chi_{\Pi v_J}^c v_J \nabla _\mu u^\mu \nonumber \\
&+& \chi_{\Pi W}^{aJ} W_\mu \nabla ^\mu \frac{\mu _J}{T} + \chi_{\Pi W}^b W_\mu \nabla ^\mu \frac{1}{T} \nonumber \\ 
&+& \chi_{\Pi W}^c W_\mu D u ^\mu + \chi_{\Pi W}^d \nabla ^\mu W_\mu \nonumber \\
&+& \chi_{\Pi V_J}^{aK} V^J_\mu \nabla ^\mu \frac{\mu_{K}}{T} + \chi_{\Pi V_J}^b V^J_\mu \nabla ^\mu \frac{1}{T} \nonumber \\
&+& \chi_{\Pi V_J}^c V^J_\mu D u ^\mu + \chi_{\Pi V_J}^d \nabla ^\mu V^J_\mu \nonumber \\
&+& \chi_{\Pi \pi} \pi _{\mu \nu} \nabla ^{\langle \mu} u^{\nu \rangle},
\end{eqnarray}
\begin{eqnarray}
v_J &=& - \zeta_{v_J v_K} D\frac{\mu_K}{T} +\zeta_{v_J \Pi} \frac{1}{T} \nabla _\mu u^\mu + \zeta_{v_J w} D\frac{1}{T} \nonumber \\
&-& \tau_{v_J v_K} D v_K - \tau_{v_J \Pi} D \Pi - \tau_{v_J w} D w \nonumber \\
&+& \chi_{v_J w}^{aK} w D\frac{\mu _K}{T} + \chi_{v_J w}^b w D \frac{1}{T} + \chi_{v_J w}^c w \nabla _\mu u^\mu \nonumber \\
&+& \chi_{v_J \Pi}^{aK} \Pi D\frac{\mu _K}{T} + \chi_{v_J \Pi}^b \Pi D \frac{1}{T} + \chi_{v_J \Pi}^c \Pi \nabla _\mu u^\mu \nonumber \\
&+& \chi_{v_J v_K}^{aL} v_K D\frac{\mu _L}{T} + \chi_{v_J v_K}^b v_K D \frac{1}{T} + \chi_{v_J v_K}^c v_K \nabla _\mu u^\mu \nonumber \\
&+& \chi_{v_J W}^{aK} W_\mu \nabla ^\mu \frac{\mu _K}{T} + \chi_{v_J W}^b W_\mu \nabla ^\mu \frac{1}{T} \nonumber \\ 
&+& \chi_{v_J W}^c W_\mu D u ^\mu + \chi_{v_J W}^d \nabla ^\mu W_\mu \nonumber \\
&+& \chi_{v_J V_K}^{aL} V^K_\mu \nabla ^\mu \frac{\mu_{L}}{T} + \chi_{v_J V_K}^b V^K_\mu \nabla ^\mu \frac{1}{T} \nonumber \\
&+& \chi_{v_J V_K}^c V^K_\mu D u ^\mu + \chi_{v_J V_K}^d \nabla ^\mu V^K_\mu \nonumber \\
&+& \chi_{v_J \pi} \pi _{\mu \nu} \nabla ^{\langle \mu} u^{\nu \rangle},
\end{eqnarray}
where the summation symbols over $J$, $K$, and $L$ are abbreviated. $\zeta$'s are the linear transport coefficients, $\tau$'s are the relaxation times, and $\chi$'s are the second order transport coefficients. The derivatives are defined as $D = u^\mu \partial_\mu$ and $\nabla^\mu = \partial^\mu - u^\mu D$ which in global equilibrium correspond to the time- and the space-derivatives. It is note-worthy that the time-like derivatives of the three dissipative quantities are present in each equation, implying that one should take an appropriate linear combination of the equations for efficient numerical estimations. One may argue that some of the thermodynamic forces can be combined using conservation laws. While such would be helpful in numerical evaluations and will be discussed in Sec.~\ref{sec:tc}, here they are kept to distinguish diagonal and off-diagonal transport coefficients and to make their physical meanings clearer. The Onsager reciprocal relations \cite{Onsager1, Onsager2} imply $\zeta_{AB} = \zeta_{BA}$ where $A,B = \Pi, w,$ and $v_J$. It should be noted that the relations are satisfied in the second order hydrodynamics considered here \cite{Monnai:2010qp} but in general can break down in the systems with external magnetic fields or rotation where reversibility of macroscopic motion is not present. At the first order, the bulk pressure is the response to the change of volume and the corresponding transport coefficient $\zeta_{\Pi \Pi}$ has to be semi-positive. The off-equilibrium corrections to the energy and conserved charges densities are the responses to the changes of inverse temperature and fugacities so $\zeta_{ww}$ and $\zeta_{v_J v_J}$ are also semi-positive. The off-diagonal linear transport coefficients can be negative but the transport coefficient matrix should be semi-positive definite.

A natural question would be whether the dissipative quantities $w$ and $v_J$ themselves, not just their thermodynamic forces, can be absorbed into the bulk pressure $\Pi$ using thermodynamic relations. In general cases this is not possible because the relaxation-type equations are dependent on initial conditions and cannot be na\"{i}vely combined. Also they appear in different terms in the conservation laws since the former is the correction to the energy/conserved charge densities while the latter is the correction to the pressure. In general cases, $e$, $P$, and $n_J$ are non-linearly related through the equation of state.
It should be noted that the dissipative quantities $w$ and $v_J$ cannot be absorbed into $e$ and $n_J$ by the frame choice either  because the number of degrees of freedom of the flow is 3 and it is used up by that of $W^\mu$ or $V_J^\mu$.
The flow is defined as $T^{\mu \nu} u_\nu = (e+w) u^\mu$ in the Landau frame and $N^\mu_J = (n_J+v_J) u^\mu$ in the Eckart frame. Here the direction of the flow $u^\mu$, or the local rest frame, is not directly changed by the presence of $w$ or $v_J$.

Stability and causality of the relativistic hydrodynamic theory is discussed in detail in Appendix A. This is important partly because the time-like derivatives in thermodynamic forces could cause instability. The causality condition is satisfied when the relaxation times $\tau_\Pi$, $\tau_w$, and $\tau_v$ are sufficiently large compared with the linear transport coefficients. The stability condition is found to be related to the positivity of entropy production.

\subsection{Linear Transport Coefficients}
\label{sec:tc}

For simplicity, I consider a system where net baryon number is the conserved charge. There are three linear transport coefficients for each dissipative quantity. One can combine them and define \textit{effective} transport coefficients as,
\begin{eqnarray}
\label{eq:effzeta}
\Pi &=& - \zeta_{\Pi} \nabla_\mu u^\mu + \mathcal{O}(\delta^2), \\
w &=& \zeta_{w} \nabla_\mu u^\mu + \mathcal{O}(\delta^2), \\
v &=& \zeta_{v} \nabla_\mu u^\mu + \mathcal{O}(\delta^2),
\end{eqnarray}
where 
\begin{eqnarray}
\begin{pmatrix}
\zeta_{\Pi} \\ \zeta_{w} \\ \zeta_{v} 
\end{pmatrix}
=
\frac{1}{T}
\mathcal{Z}
\mathbf{a} ,
\end{eqnarray}
with the transport coefficient matrix
\begin{eqnarray}
\mathcal{Z} = 
\begin{pmatrix}
\zeta_{\Pi \Pi} & \zeta_{\Pi w}& \zeta_{\Pi v} \\
\zeta_{w \Pi} & \zeta_{w w} & \zeta_{w v} \\
\zeta_{v \Pi} & \zeta_{v w} & \zeta_{v v}
\end{pmatrix} , \label{eq:tmat}
\end{eqnarray}
and an auxiliary vector
\begin{eqnarray}
\label{eq:veca}
\mathbf{a} = 
\begin{pmatrix}
1 \\ ( \frac{\partial P}{\partial e})_{n_B} \\ ( \frac{\partial P}{\partial n_B})_{e}
\end{pmatrix} ,
\end{eqnarray}
using the hydrodynamic identities that are derived from the conservation laws and the Gibbs-Duhem relation $dP=sdT+n_Bd\mu_B$ which represents the first law of thermodynamics \cite{Hosoya:1983xm},
\begin{eqnarray}
D\frac{1}{T} 
&=& \bigg( \frac{\partial P}{\partial e} \bigg)_{n_B} \frac{1}{T} \nabla_\mu u^\mu + \mathcal{O}(\delta^2) , \\
D\frac{\mu_B}{T} 
&=& - \bigg( \frac{\partial P}{\partial n_B} \bigg)_{e} \frac{1}{T} \nabla_\mu u^\mu + \mathcal{O}(\delta^2) .
\end{eqnarray}
The relations are truncated at the first order because it is sufficient for the discussion of the linear transport coefficients here.

So far there are very few quantitative studies on the coefficients for $w$ and $v$, let alone the cross coefficients, in a QCD system. Here I approach the issue in the following way. A gauge-gravity correspondence analysis suggests the lower limit of the bulk viscosity is
$\zeta_{\Pi} = 2(1/3 -c_s^2) \eta$ \cite{Buchel:2007mf}, 
which can be recovered when $\zeta_{\Pi w} = -3 \zeta_{\Pi \Pi}$ and $\zeta_{\Pi v} = [n_B/(e+P)] \zeta_{\Pi w}$ where $\eta$ is the shear viscosity. Note that the sound velocity is expressed as
\begin{equation}
c_s^2 = \bigg( \frac{\partial P}{\partial e} \bigg)_{n_B} + \frac{n_B}{e+P} \bigg( \frac{\partial P}{\partial n_B} \bigg)_e .
\label{eq:cs2}
\end{equation}
Keeping in mind that the conjectured lower boundary is $\eta/s = 1/4\pi$ \cite{Kovtun:2004de} and replacing the entropy density with the enthalpy over temperature at finite density, one may use for demonstration the parametrizations
\begin{eqnarray}
\zeta_{\Pi} &=& C_\Pi \bigg(\frac{1}{3} -c_s^2\bigg) \frac{e+P}{4\pi T}, \label{eq:CPi} \\
\zeta_{w} &=& C_w \bigg(\frac{1}{3} -c_s^2\bigg) \frac{e+P}{4\pi T} , \label{eq:Cw} \\
\zeta_{v} &=& C_v \bigg(\frac{1}{3} -c_s^2\bigg) \frac{n_B}{4\pi T} , \label{eq:Cv} 
\end{eqnarray}
where $C_\Pi$, $C_w$ and $C_v$ are dimensionless factors. 
Since the entropy production has to be semi-positive, in the absence of vector and tensor dissipative currents the transport coefficients are subject to the following constraint at the linear order:
\begin{eqnarray}
\partial_\mu s^\mu
&=& w D\frac{1}{T} - \Pi \frac{1}{T} \nabla_\mu u^\mu - v D\frac{\mu_B}{T} \nonumber \\
&=& \bigg[ \zeta_w \bigg( \frac{\partial P}{\partial e} \bigg)_{n_B} + \zeta_\Pi + \zeta_v \bigg( \frac{\partial P}{\partial n_B} \bigg)_{e}\bigg] \nonumber \\
&\times& \frac{1}{T} (\nabla_\mu u^\mu)^2 \geq 0 , \label{eq:posen}
\end{eqnarray}
where $s^\mu$ is the entropy current. This is equivalent to the condition that the transport coefficient matrix (\ref{eq:tmat}) is semi-positive definite because the sum of effective transport coefficients can be expressed in a quadratic form as
\begin{eqnarray}
Z &\equiv& \zeta_w \bigg( \frac{\partial P}{\partial e} \bigg)_{n_B} + \zeta_\Pi + \zeta_v \bigg( \frac{\partial P}{\partial n_B} \bigg)_{e} \nonumber \\
&=& \mathbf{a}^T \mathcal{Z} \mathbf{a} , \label{eq:posen2}
\end{eqnarray}
using Eqs.~(\ref{eq:effzeta})-(\ref{eq:veca}).
It can also be shown that this condition is equivalent to the hydrodynamic stability condition obtained through linear perturbation analyses at the first order (Appendix~\ref{sec:sta}). It should be noted that the effective transport coefficients $\zeta_w$ and $\zeta_v$ can be negative unlike the orthogonal ones $\zeta_{ww}$ and $\zeta_{vv}$. Several values of the factors are used to demonstrate the interplay of those dissipative processes in Sec.~\ref{sec:num}. 

\section{Numerical analyses of heavy-ion collisions}
\label{sec:num}
\vspace*{-2mm}

In this section, I study the effects of the off-equilibrium corrections to the energy and the baryon number densities on heavy-ion observables in numerical estimations. The (2+1)-dimensional boost-invariant hydrodynamic model is employed \cite{Monnai:2014kqa}. The equation of state is based on hadron resonance gas model and lattice QCD estimations \cite{Bazavov:2014pvz, Bazavov:2012jq, Ding:2015fca}. The Monte-Carlo Glauber model is used to construct initial conditions. For the demonstrative nature of the present study, the initial conditions are smoothed over by taking average over events. The normalization of energy distribution is determined so that the identified particle spectra of Au-Au collisions at $\sqrt{s_{NN}} = 200$ GeV \cite{Adler:2003cb} are reproduced after resonance decays in the most central collisions. The baryon number distribution is normalized so that the $s/n_B$ ratio is fixed to 420 \cite{Gunther:2016vcp}. The events of 20-30\% centralities are considered. The initial time for hydrodynamic evolution is $\tau_\mathrm{th} = 0.4$ fm/$c$. The resonance decays are treated as in Ref.~\cite{Sollfrank:1990qz}. Full comparison to the experimental data and detailed tuning of the transport coefficients are beyond the scope of this study and will be discussed elsewhere.

Bulk viscosity is implemented but shear viscosity and baryon diffusion are not because Curie's theorem implies that only scalar dissipative currents are mixed at the linear order in an isotropic system. Since the aforementioned gauge-gravity correspondence approach conjectures that lower boundary is $C_\Pi = 2$ \cite{Buchel:2007mf}, the first order transport coefficients are parametrically chosen as in Eqs.~(\ref{eq:CPi})-(\ref{eq:Cv}) with $(C_\Pi, C_w, C_v) = (2,0,0), (2,\pm6,0),$ and $(2,\pm6, \pm6)$. Note that negative effective transport coefficients are allowed as long as they satisfy the condition of semi-positive entropy production (\ref{eq:posen}). For simplicity, the second order transport coefficients are set to vanishing, \textit{i.e.}, $\chi_{AB} = \tau_{AB} = 0$ ($A,B = \Pi, w,$ and $v$) except for the diagonal relaxation times. There has been very few study on those relaxation times so they are simply assumed to be $\tau_\Pi = \tau_w = \tau_v = \tau_R$ where $\tau_R$ is taken from the bulk viscous relaxation time in Ref.~\cite{Natsuume:2007ty}.

It has been known that the effect of off-equilibrium corrections appears in particle spectra not only through the modification of flow profile but also through the distorted phase space distribution at freeze-out used in the Cooper-Frye formula \cite{Cooper:1974mv,Teaney:2003kp},
\begin{eqnarray}
\frac{dN_i}{d^2p_T dy_p} = \frac{g_i}{(2\pi)^3} \int_\Sigma p_i^\mu (f^0_i + \delta f_i), \label{eq:CF}
\end{eqnarray}
where $p_T$ is the transverse momentum, $y_p$ is the rapidity, $g_i$ is the degeneracy, $f_i^0$ is the equilibrium distribution, and $\delta f_i$ is the off-equilibrium correction to the distribution. The freeze-out energy density $e_f = 0.14$ GeV is used for determining the hypersurface $\Sigma$. It roughly translates into $T = 0.14$ GeV at the vanishing chemical potential. The form of $\delta f_i$ is based on Ref.~\cite{Monnai:2009ad, Monnai:2010qp}. The details are summarized in Appendix~\ref{sec:deltaf}.

\subsection{Hydrodynamic Evolution}

\begin{figure}[tb]
\includegraphics[width=3.4in]{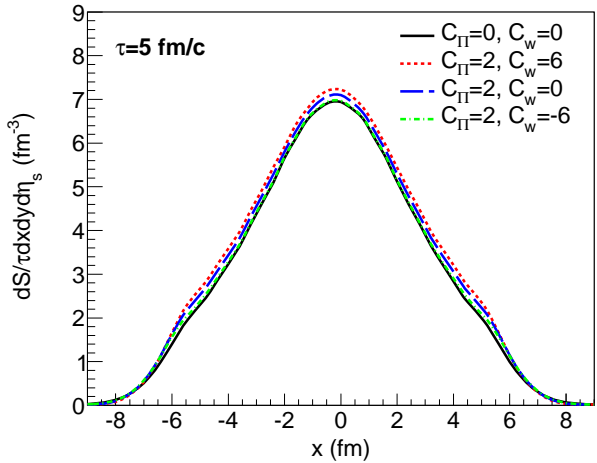}
\caption{(Color online) The entropy distribution at $y = \eta_s = 0$ at $\tau = 5$ fm/$c$ for the ideal fluid (solid line) and for the dissipative fluids with $(C_\Pi, C_w) = (2,6)$ (dotted line), $(2,0)$ (dashed line), and $(2,-6)$ (dash-dotted line). $C_v$ is set to vanishing.}
\label{fig:1}
\end{figure}

First, the space time evolution of entropy distribution is investigated. The entropy density in an off-equilibrium system is given as $s = s \cdot u = s_0 + (w - \mu_B v)/T$ neglecting the higher order corrections. Note that $\Pi$ does not appear explicitly in the expression. Figure \ref{fig:1} shows the entropy distribution on the $y = \eta_s = 0$ plane at $\tau = 5$ fm/$c$ with $(C_\Pi, C_w, C_v) = (2,0,0), (2,\pm6,0)$. The initial conditions are the same for all cases. One can see that the pure effect of bulk viscosity enhances the entropy distribution as is well known. The positive $w$, which follows from the positive $C_w$, also enhances the distribution through entropy production, as expected from Eq.~(\ref{eq:posen}). The negative $w$ reduces the entropy distribution, but detailed analyses shows that the overall entropy production is still positive. With this specific choice of transport coefficients, the dissipative effects are almost cancelled because $(\partial P/\partial e)_{n_B} \sim c_s^2 \sim 1/3$ at small baryon number density in Eq.~(\ref{eq:posen}). The cancellation is slightly weak in the peripheral regions in Fig. \ref{fig:1} because the QCD sound velocity is slowed down near the crossover.

The effect of the correction to the baryon number density is also estimated numerically using the parameter sets $(C_\Pi, C_w, C_v) = (2,\pm6, \pm6)$. It is found to be negligible at the top RHIC energies because of the small baryon density in the system. Its effect at lower beam energies is an interesting topic and and is left for future studies. 

\subsection{$p_T$ Spectra}

\begin{figure}[tb]
\includegraphics[width=3.4in]{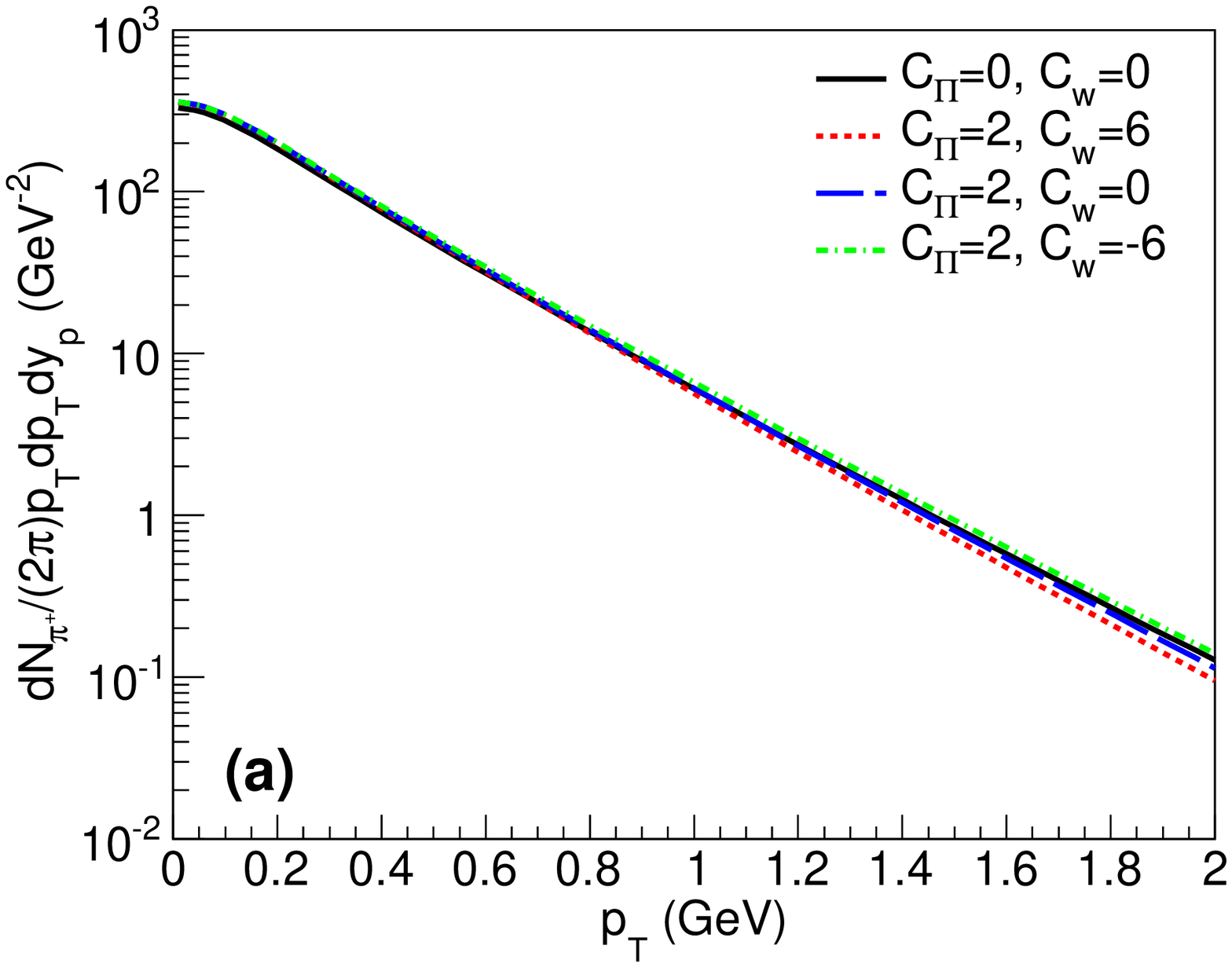}
\includegraphics[width=3.4in]{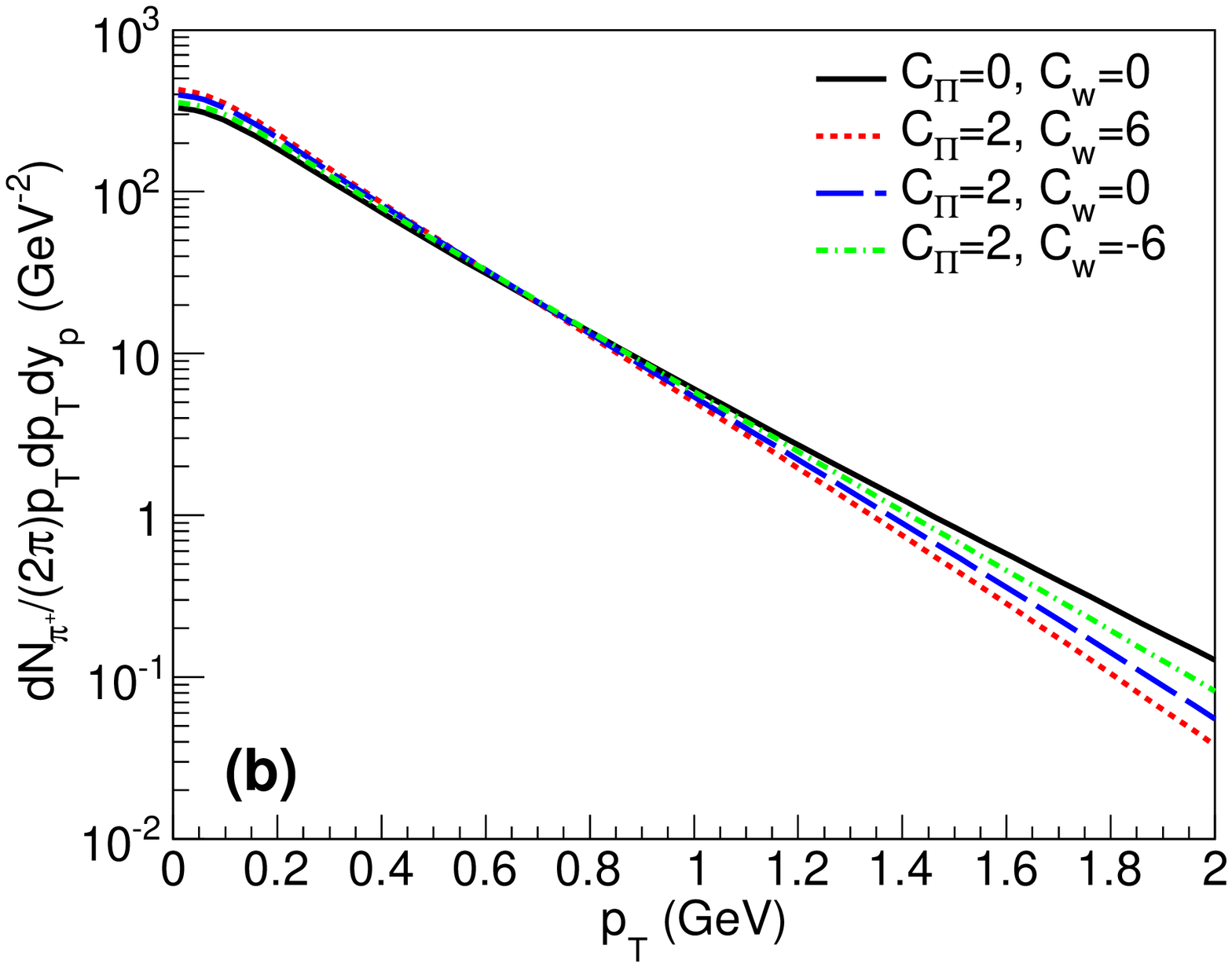}
\caption{(Color online) $p_T$ spectra of $\pi^+$ (a) without and (b) with $\delta f$ correction for the ideal fluid (solid line) and for the dissipative fluids with $(C_\Pi, C_w) = (2,6)$ (dotted line), $(2,0)$ (dashed line), and $(2,-6)$ (dash-dotted line). $C_v$ is set to vanishing.}
\label{fig:2}
\end{figure}

The particle spectra of positive pions at midrapidity after resonance decays are shown without and with the $\delta f$ corrections at freeze-out (\ref{eq:CF}) in Fig.~\ref{fig:2} (a) and (b), respectively. The ideal hydrodynamic result is compared with the dissipative hydrodynamic ones with the bulk viscosity and energy density correction. The coefficients $(C_\Pi, C_w, C_v) = (2,0,0), (2,\pm6,0), (2,\pm6, \pm6)$ are used for the numerical estimations but the finite $C_v$ results are not shown as its effect is negligible for the current choice of parameters. The transverse momentum range $0 \leq p_T \leq 2$~GeV is shown because the results would not be reliable when the effects of the $\delta f$ correction, which is based on momentum expansion, is too large at higher $p_T$. Here the correction is truncated if $|\delta f / f_0| > 0.9$ to avoid overestimation. 

The results in Fig.~\ref{fig:2} (a) shows that the dissipative effects are small without the $\delta f$ correction, but more quantitative analyses indicate that bulk viscosity enhances the particle number and reduce the mean $p_T$ via modification of the hydrodynamic flow. It can be shown, on the other hand, that the positive energy density correction with $C_w = 6$ further reduces the mean $p_T$ spectrum by 4.4\% but reduces the number of particles by 2.5\%. This apparent discrepancy between the reduced particle number and positive entropy production is caused by the lack of the $\delta f$ correction at freeze-out. Figure~\ref{fig:2} (b) shows that the positive energy density correction is shown to reduce mean $p_T$ by 1.6\% and increase the number of particles by 2\%, which is qualitatively similar to the bulk viscosity, when the $\delta f$ corrections are properly handled.
The trend is reversed for the negative energy density correction with $C_w = -6$ and partial cancellation of the effects of the two types of dissipative quantities can be found, as has been the case in Fig.~\ref{fig:1}.

The entropy production caused by the dissipative processes is checked to be positive. It should be noted that $\delta f$ correction must be taken into account to correctly understand the entropy production caused by the correction to the energy density $w$ because it is non-vanishing at the linear order perturbation of $s^\mu$ unlike the bulk pressure $\Pi$.
Also the increase of the entropy does not necessarily mean the increase of the particle number because if the modified distribution tends to produce more heavier particles and less light ones, the total particle number could be lowered while the entropy is increased. The effects of resonance decays thus become more important.

$p_T$ spectra for kaons and protons are also calculated and the dissipative corrections are found to have similar effects on the particle spectra to those on the pionic ones. The numbers of pions, kaons and protons are all enhanced after resonance decays. 

\subsection{Elliptic Flow}

Finally, I study the hadronic elliptic flow $v_2$. The differential elliptic flow coefficient is estimated as 
\begin{eqnarray}
v_2(p_T,y_p) = \frac{\int d\phi_p \cos[2(\phi_p-\Psi)] \frac{dN}{d\phi_p p_T dp_T dy_p}}{\int d\phi_p \frac{dN}{d\phi_p p_T dp_T dy_p}}.
\end{eqnarray}
where $\phi_p$ is the azimuthal angle in momentum space and $\Psi$ is the event plane angle. 

\begin{figure}[tb]
\includegraphics[width=3.4in]{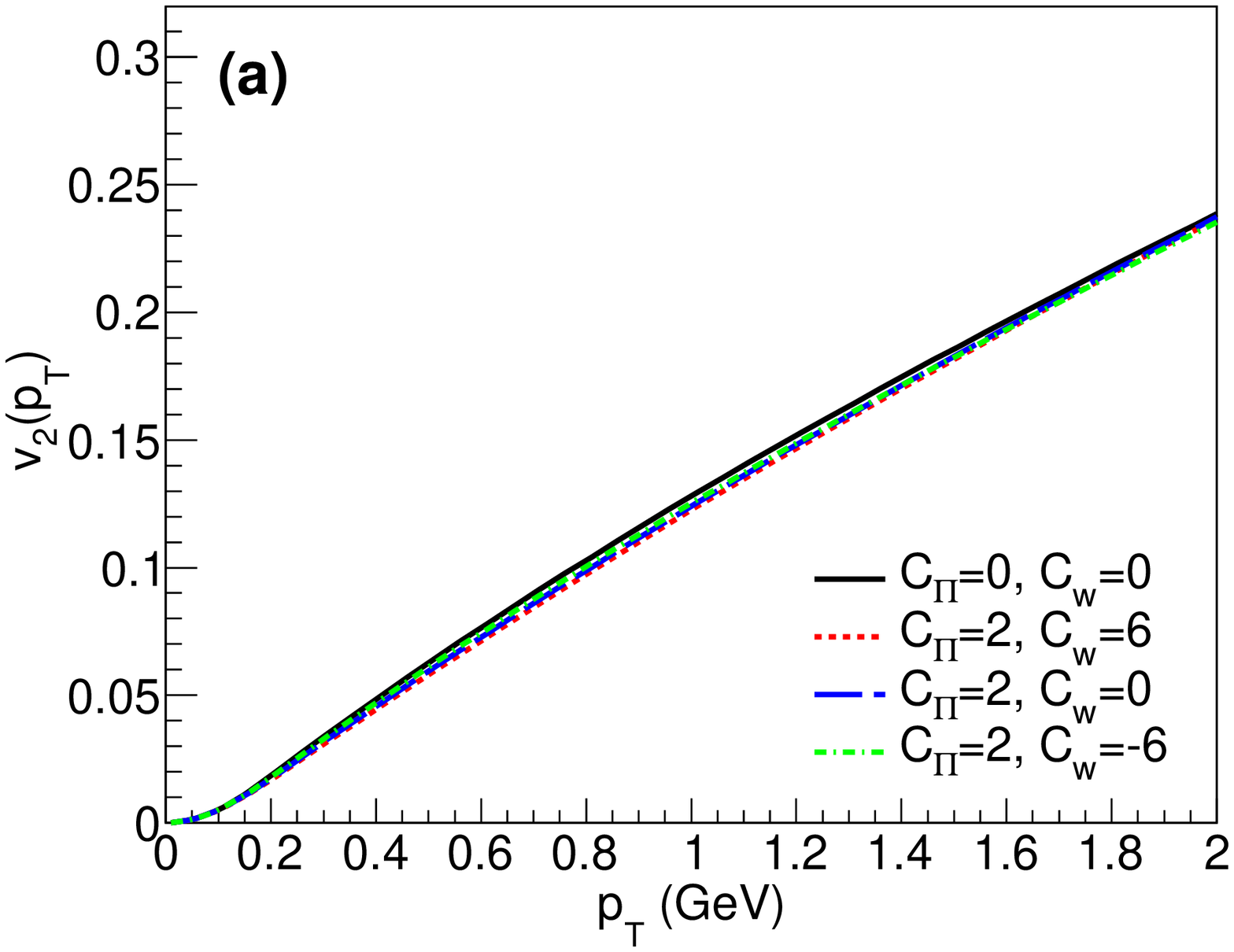}
\includegraphics[width=3.4in]{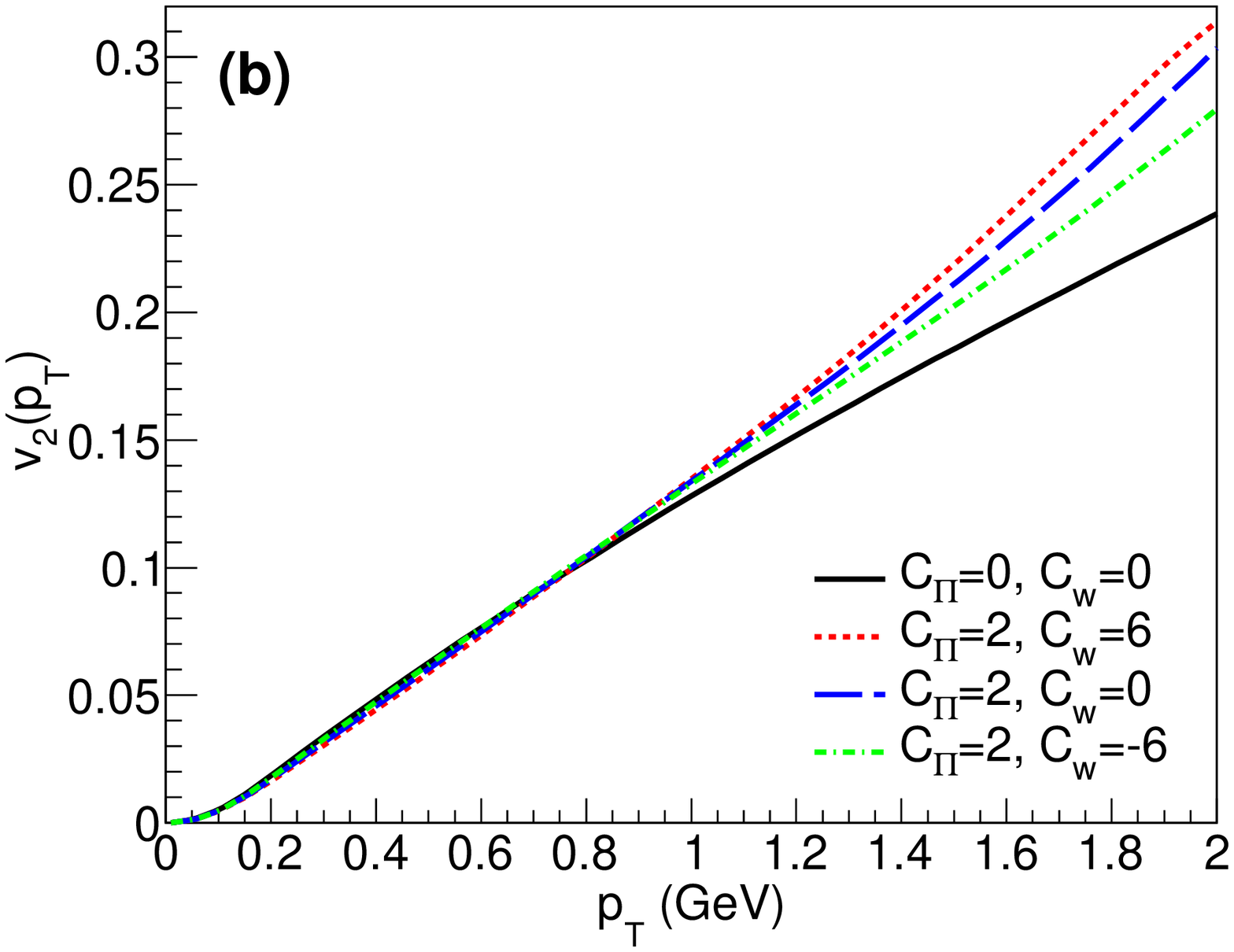}
\caption{(Color online) Differential elliptic flow $v_2(p_T)$ of $\pi^+$ (a) without and (b) with $\delta f$ correction for the ideal fluid (solid line) and for the dissipative fluids with $(C_\Pi, C_w) = (2,6)$ (dotted line), $(2,0)$ (dashed line), and $(2,-6)$ (dash-dotted line). $C_v$ is set to vanishing.}
\label{fig:3}
\end{figure}

$v_2$ of positive pions at $y_p=0$ are shown in Fig.~\ref{fig:3} for the ideal and the dissipative cases. Fig.~\ref{fig:3}~(a) shows that the elliptic flow is not much effected but is slightly reduced by the bulk viscosity and also by the correction to the energy density for the positive $C_w$ without the $\delta f$ correction. Figure~\ref{fig:3}~(b) illustrates that the off-equilibrium distribution enhances the elliptic flow for both cases above around $p_T \sim 0.7$ GeV. 
The enhancement of $v_2$ by the correction to the energy density and the bulk viscosity in the full estimation may be attributed to the reduction of mean $p_T$ \cite{Monnai:2009ad, Hirano:2005wx}. Further numerical estimations show that the enhancement effect is also found in kaons and protons. The effect of the energy density correction with the negative $C_w$ again is found to partially cancel with that of the bulk viscosity.

\section{Discussion and Conclusions}
\label{sec:disconc}
\vspace*{-2mm}

I have investigated the possible off-equilibrium corrections to the energy and the conserved charge densities, which follow from the straight forward derivation of Israel-Stewart-type relativistic dissipative hydrodynamic equations. They are conventionally put to vanishing for the Landau matching conditions, but they could be modified by redefining the local equilibrium. Those dissipative quantities are described by constitutive equations similar to that of the bulk viscosity, but they cannot be mathematically merged as they appear in different terms of the conservation laws. 
The transport coefficients are constrained by the condition of positive entropy production. The condition is also shown to be equivalent to the hydrodynamic stability condition at the first order. 

Numerical estimations are performed to understand the phenomenological consequences of the corrections to the energy and baryon number densities in ultrarelativistic nuclear collisions with a set of transport coefficients motivated by the conjectured lower boundary for bulk viscosity \cite{Buchel:2007mf}. I have developed a (2+1)-dimensional dissipative hydrodynamic code that can estimate the newly introduced dissipative currents. Instability is not observed during hydrodynamic evolution for the transport coefficients that satisfy the condition of positive entropy production. The entropy distribution is found to be enhanced by the correction to the energy density with a positive effective coefficient $\zeta_w > 0$. The first order dissipative correction to the entropy density has to be explicitly considered. It is found that the effect of the energy density correction with a negative effective coefficient $\zeta_w < 0$ and that of the bulk viscosity can partially cancel out. The effect of the correction to the baryon number density is also estimated and is found to be small at $\sqrt{s_{NN}} = 200$ GeV for the investigated values of $\zeta_v$. 

The particle spectra of positive pions, kaons and protons are then estimated. It is important to take into consideration the off-equilibrium corrections to the phase-space distribution at freeze-out even for the qualitative understanding of those observables because the energy density correction, unlike the bulk viscosity, appears at the first order in the expansion of the entropy density around equilibrium. Thus one loses entropy by neglecting the distortion of distribution. Numerical estimation indicates that the number of particles is enhanced and the mean $p_T$ is reduced by the correction to the energy density when the $\delta f$ correction and hadronic decays are taken into account. It is note-worthy that there is an on-going debate on the form of the off-equilibrium distribution \cite{Denicol:2009am, Monnai:2009ad, Bozek:2009dw, Dusling:2011fd, Noronha-Hostler:2013gga, Molnar:2014fva}. 

The differential elliptic flow is estimated and is found to be increased by the energy density correction except near very small $p_T$. This is similar to the case of bulk viscosity. It implies that the existence of off-equilibrium correction to the energy density would interfere with the extraction of bulk viscous coefficient from the experimental data, because its would be difficult to distinguish the effects of the two types of dissipative currents.

Future prospects include numerical estimation of the effects of dissipative correction to the baryon number density in the systems with higher baryon chemical potential for the Beam Energy Scan program at RHIC. Also it is important to theoretically establish the formulation of causal dissipative relativistic hydrodynamics to understand what those types of dissipative quantities mean and if they should be allowed. It has been shown that they would not cause hydrodynamic nor numerical instabilities in this work, but further investigation would be required. Finally, it might be necessary to find experimental observables that are sensitive to the difference between the effects of bulk viscosity and energy density correction for better understanding of the transport phenomena in a QCD matter. Electromagnetic probes could be used for the purpose since their emission rates are expected to be dependent on the dissipative corrections to the phase-space distribution functions.

\begin{acknowledgments}
The author is grateful for the valuable discussion with Jean-Yves Ollitrault. 
The work of A.M. is supported by JSPS Overseas Research Fellowships.
\end{acknowledgments}

\appendix

\section{HYDRODYNAMIC STABILITY AND CAUSALITY}
\label{sec:sta}

I investigate hydrodynamic stability and causality of the system with off-equilibrium corrections following the method of Ref.~\cite{Hiscock:1985zz,Denicol:2008rj,Pu:2009fj}. Here I focus on the scalar dissipative quantities in Landau frame and set aside the energy dissipation $W^\mu$, the charge diffusion $V^\mu$, and the shear stress tensor $\pi^{\mu \nu}$. 
The macroscopic variables can be separated into global equilibrium and perturbation components as $Q = Q_\mathrm{eq} + \delta Q$.
The dissipative currents are vanishing in global equilibrium so the terms that involve $w_\mathrm{eq}$, $\Pi_\mathrm{eq}$, and $v_\mathrm{eq}$ vanish. 
The perturbation of the conserving quantities up to the first order in $\delta Q$ then reads
\begin{eqnarray}
\delta T^{\mu \nu} &=& (e^\mathrm{eq}+P^\mathrm{eq}) (\delta u^\mu u_\mathrm{eq}^\nu + u_\mathrm{eq}^\mu \delta u^\nu) \nonumber \\
&+& (\delta e + \delta w) u_\mathrm{eq}^\mu u_\mathrm{eq}^\nu - (\delta P + \delta \Pi) g^{\mu \nu}, \\
\delta N^\mu &=& n^\mathrm{eq} \delta u^\mu + (\delta n + \delta v ) u_\mathrm{eq}^\mu .
\end{eqnarray}
Similarly, one can consider the perturbation of the second-order constitutive equations. They are expressed as
\begin{eqnarray}
\delta w &=& \zeta_{w} \nabla_\mu \delta u^\mu - \tau_w D \delta w - \tau_{w\Pi} D \delta \Pi - \tau_{wv} D \delta v , \\
\delta \Pi &=& - \zeta_{\Pi} \nabla_\mu \delta u^\mu - \tau_{w\Pi} D \delta w - \tau_{\Pi} D \delta \Pi - \tau_{\Pi v} D \delta v, \nonumber \\ \\
\delta v &=& \zeta_{w} \nabla_\mu \delta u^\mu - \tau_{wv} D \delta w - \tau_{\Pi v} D \delta \Pi - \tau_{v} D \delta v .
\end{eqnarray}

Here I consider a plane-wave deviation from equilibrium $\delta Q = \delta \bar{Q} e^{i(\omega t - kx)}$ where $\delta \bar{Q}$ is independent of the space-time coordinates. 
Keeping in mind that $u^\mu_\mathrm{eq} = (1,0,0,0)$, the perturbed equations of motion involving the target dissipative quantities can be expressed in a matrix form as:
\begin{eqnarray}
\label{eq:longitudinal}
\mathcal{M}_{xx}
\begin{pmatrix}
\delta e\\
\delta n\\
\delta u^x \\
\delta w \\
\delta \Pi \\
\delta v 
\end{pmatrix}
&=&0 ,
\end{eqnarray}
where the matrix is explicitly expressed as
\begin{widetext}
\begin{eqnarray}
\label{eq:Mxx}
\mathcal{M}_{xx} = 
\begin{pmatrix}
i \omega&0&-ik(e+P)&i \omega&0&0 \\ 
-ik( \frac{\partial P}{\partial e})_n&-ik( \frac{\partial P}{\partial n})_e&i \omega(e+P)&0&-ik&0 \\ 
0&i \omega&- ikn&0&0&i \omega \\ 
0&0&ik\zeta_{w}&1+i \omega \tau_w&i \omega \tau_{\Pi w}&i \omega \tau_{w v} \\ 
0&0&-ik\zeta_{\Pi}&i \omega \tau_{\Pi w}&1+i \omega \tau_\Pi&i \omega \tau_{\Pi v} \\ 
0&0&ik\zeta_{v}&i \omega \tau_{w v}&i \omega \tau_{\Pi v}&1+i \omega \tau_v \\ 
\end{pmatrix} .
\end{eqnarray}
\end{widetext}
The subscript for global equilibrium is abbreviated. Here the equations for the longitudinal modes are considered because the scalar dissipative currents do not appear in those for the transverse modes.

The equations have non-trivial solutions if the determinant of the matrix is vanishing, $\mathrm{det}(\mathcal{M}_{xx}) = 0$. The system is unstable when the imaginary part of $\omega$ is negative since that is the indication of a growing mode. Thus the stability condition can be written here as
\begin{eqnarray}
\label{eq:stability}
\mathrm{Im} (\omega) \geq 0.
\end{eqnarray}
The causality condition, on the other hand, can be expressed that the group velocity does not exceed unity:
\begin{eqnarray}
\label{eq:causality}
\bigg| \frac{\partial \mathrm{Re} (\omega)}{\partial k} \bigg| &\leq& 1 .
\end{eqnarray}

\subsection{First Order Theory}

It is generally very complicated to solve those equations completely analytically.
In the first order limit where $\tau_w, \tau_\Pi, \tau_v, \tau_{\Pi w}, \tau_{\Pi v}, \tau_{wv} \to 0$, the exact solutions are $\omega = 0$ and 
\begin{eqnarray}
\label{eq:Glin}
\omega &=& \frac{i}{2(e+P)}[ k^2 Z \pm \sqrt{ k^4 Z^2 - 4k^2 (e+P)^2 c_s^2} ] , 
\end{eqnarray}
where
\begin{equation}
Z = \bigg( \frac{\partial P}{\partial e} \bigg)_n \zeta_w + \zeta_\Pi + \bigg( \frac{\partial P}{\partial n} \bigg)_e \zeta_v .
\label{eq:Z}
\end{equation}
Here the expression of the sound velocity (\ref{eq:cs2}) is used. The imaginary part of Eq.~(\ref{eq:Glin}) is semi-positive when $Z \geq 0$. It is note-worthy that this condition is equivalent to the condition of positive entropy production (\ref{eq:posen}). 
On the other hand, the group velocity can exceed unity for sufficiently large $k$, implying the necessity of introducing second order theory. The details can be found in Sec.~\ref{sec:sot}.

\subsection{Second Order Theory}
\label{sec:sot}

The dispersion relation for the second order theory can be written from $\mathrm{det}(\mathcal{M}_{xx}) = 0$ as, aside from the trivial $\omega = 0$, 
\begin{eqnarray}
\label{eq:diss}
\omega^2 - c_s^2 k^2 &=& \frac{i\omega k^2}{(e+P)} \nonumber \\
&\times&\frac{\zeta_\Pi T_w T_v + ( \frac{\partial P}{\partial e} )_n \zeta_w T_\Pi T_v + ( \frac{\partial P}{\partial n} )_e \zeta_v T_\Pi T_w}{T_\Pi T_w T_v}, \nonumber \\
\end{eqnarray}
where
\begin{eqnarray}
\label{eq:diss2}
T_\Pi &=& 1 + i \tau_\Pi \omega , \\
T_w &=& 1 + i \tau_w \omega , \\
T_v &=& 1 + i \tau_v \omega ,
\end{eqnarray}
when the cross relaxation times $\tau_{\Pi w}$, $\tau_{\Pi v}$, and $\tau_{wv}$ are neglected.
Since there is no simple analytic solution to the quintic equation, I first focus on the off-equilibrium correction to the energy density. The equation reduces to 
\begin{eqnarray}
\label{eq:diss3}
\omega^2 - c_s^2 k^2 = \frac{i\omega k^2 ( \frac{\partial P}{\partial e} )_n \zeta_w}{(e+P)(1 + i \tau_w \omega)}.
\end{eqnarray}
This is the same dispersion relation as the one for bulk viscosity with $\zeta_\Pi$ and $\tau_\Pi$ \cite{Denicol:2008rj,Pu:2009fj} substituted by $( \frac{\partial P}{\partial e} )_n \zeta_w$ and $\tau_w$, respectively. The exact solutions can be obtained analogously. Physical insights can be obtained by considering asymptotic forms for large and small $k$. For large $k$, they read
\begin{eqnarray}
\label{eq:diss4}
\omega &=& \pm k \sqrt{c_s^2 + \frac{( \frac{\partial P}{\partial e} )_n \zeta_w}{\tau_w (e+P)}} \nonumber \\
&+& \frac{i( \frac{\partial P}{\partial e} )_n\zeta_w}{2\tau_w [( \frac{\partial P}{\partial e} )_n \zeta_w+c_s^2 \tau_w (e+P)]}, 
\end{eqnarray}
which are propagating modes, and 
\begin{eqnarray}
\label{eq:diss5}
\omega &=& \frac{ic_s^2 (e+P)}{( \frac{\partial P}{\partial e} )_n \zeta_w+c_s^2 \tau_w (e+P)} ,
\end{eqnarray}
which is a non-propagating mode. On the other hand, for small $k$ the propagating modes are
\begin{eqnarray}
\label{eq:diss6}
\omega &=& \pm k c_s + \frac{i ( \frac{\partial P}{\partial e} )_n \zeta_w k^2}{2(e+P)} ,
\end{eqnarray}
and the non-propagating mode is
\begin{eqnarray}
\label{eq:diss7}
\omega &=& \frac{i}{\tau_w}.
\end{eqnarray}

The maximum group velocity of the propagating mode is given as
\begin{eqnarray}
\label{eq:diss8}
\bigg| \frac{\partial \mathrm{Re} (\omega)}{\partial k} \bigg| &=& \sqrt{c_s^2 + \frac{(\frac{\partial P}{\partial e} )_n \zeta_w}{\tau_w (e+P)}} ,
\end{eqnarray}
since $(\frac{\partial P}{\partial e} )_n \zeta_w \geq 0$ from the law of increasing entropy. The condition of causality can be satisfied for a sufficiently large relaxation time $\tau_w \geq  (\frac{\partial P}{\partial e} )_n \zeta_w/[(e+P)(1-c_s^2)]$. The velocity diverges in the first order limit $\tau_w \to 0$. It is note-worthy that there is an argument that the region of superluminal propagation can already be out of the range of hydrodynamic applicability \cite{Van:2007pw}.

The imaginary part of $\omega$ stays semi-positive, \textit{i.e.}, the system is stable if $(\frac{\partial P}{\partial e} )_n \zeta_w \geq 0$, which is also the condition of the positive entropy production.

Next the effect of off-equilibrium correction to the conserved charge density is investigated. The non-trivial dispersion relation leads to 
\begin{eqnarray}
\label{eq:diss9}
\omega^2 - c_s^2 k^2 = \frac{i\omega k^2 ( \frac{\partial P}{\partial n} )_e\zeta_n}{(e+P)(1 + i \tau_v \omega)}.
\end{eqnarray}
As one can immediately see, this is a substitution of $( \frac{\partial P}{\partial e} )_n \zeta_w$ and $\tau_w$ by $( \frac{\partial P}{\partial n} )_e \zeta_v$ and $\tau_v$, respectively. The group velocity is 
\begin{eqnarray}
\label{eq:diss10}
\bigg| \frac{\partial \mathrm{Re} (\omega)}{\partial k} \bigg| &=& \sqrt{c_s^2 + \frac{(\frac{\partial P}{\partial n} )_e \zeta_v}{\tau_v (e+P)}} ,
\end{eqnarray}
which should again be smaller than unity. The stability condition is given as $(\frac{\partial P}{\partial n} )_e \zeta_v \geq 0$.

Finally, when the three dissipative processes are present, the expression becomes much complicated. If one employs the simple relaxation time approximation $\tau_R = \tau_\Pi = \tau_w = \tau_v$, then the dispersion relation is 
\begin{eqnarray}
\label{eq:diss11}
\omega^2 - c_s^2 k^2 = \frac{i\omega k^2 Z}{(e+P)(1 + i \tau_R \omega)},
\end{eqnarray}
which leads to 
\begin{eqnarray}
\label{eq:diss12}
\bigg| \frac{\partial \mathrm{Re} (\omega)}{\partial k} \bigg| &=& \sqrt{c_s^2 + \frac{Z}{\tau_R (e+P)}} .
\end{eqnarray}
The stability condition is satisfied when $Z \geq 0$. The results implies that the condition of positive entropy production is closely related with the hydrodynamic stability of the system.

\section{OFF-EQUILIBRIUM PHASE-SPACE DISTRIBUTION}
\label{sec:deltaf}

The dissipative corrections to the distribution $\delta f^i$ of the $i$-th particle species can be determined using the Grad moment method and the self-consistency condition that the modified distribution leads to the correct off-equilibrium energy-momentum tensor and the net baryon number in kinetic theory \cite{Israel:1979wp, Monnai:2009ad, Monnai:2010qp}. The off-equilibrium component of the phase-space distribution is
\begin{eqnarray}
\label{eq:df}
\delta f^i &=& - f_0^i (1\pm f_0^i) \{ b_i p_i^\mu (D_w w + D_\Pi \Pi + D_v v) u_\mu \nonumber \\
&+& p_i^\mu p_i^\nu [(B_w w + B_\Pi \Pi + B_v v) \Delta _{\mu \nu} \nonumber \\
&+& (\tilde{B}_w w + \tilde{B}_\Pi \Pi + \tilde{B}_v v) u_\mu u_\nu ] \} + \mathcal{O}(\delta^2) ,
\end{eqnarray}
where the sign is positive for bosons and negative for fermions.
The explicit expressions of the coefficients are
\begin{eqnarray}
D_\Pi = 3 (J_{40} J^B_{31} - J_{41} J^B_{30}) \mathcal{J}_3^{-1} ,\\
B_\Pi = (J^B_{30} J^B_{30} - J_{40} J^{BB}_{20}) \mathcal{J}_3^{-1} ,\\
\tilde{B}_\Pi = 3 (J_{41} J^{BB}_{20} - J^B_{30} J^B_{31}) \mathcal{J}_3^{-1} , \\
D_w = (3 J_{41} J^B_{31} - 5 J_{42} J^B_{30}) \mathcal{J}_3^{-1} ,\\
B_w = (J^B_{30} J^B_{31} - J_{41} J^{BB}_{20}) \mathcal{J}_3^{-1} ,\\
\tilde{B}_w = (5 J_{42} J^{BB}_{20} - 3 J^B_{31} J^B_{31}) \mathcal{J}_3^{-1} , \\
D_v = (5 J_{42} J_{40} - 3 J_{41} J_{41}) \mathcal{J}_3^{-1} ,\\
B_v = (J_{41} J^{B}_{30} - J_{40} J^B_{31}) \mathcal{J}_3^{-1} ,\\
\tilde{B}_v = (3 J_{41} J^B_{31} - 5 J_{42} J^B_{30}) \mathcal{J}_3^{-1} ,
\end{eqnarray}
where 
\begin{eqnarray}
\mathcal{J}_3 &=& 5 J_{42} J^B_{30} J^B_{30} + 3 J^B_{31} J_{40} J^B_{31} + 3 J_{41} J_{41} J^{BB}_{20} \nonumber \\
&-& 3 J^B_{31} J_{41} J^B_{30} - 3 J_{41} J^B_{30} J^B_{31} - 5 J_{42} J_{40} J^{BB}_{20}
\label{eq:detJ3} .
\end{eqnarray}
The index $B$ denotes the presence of baryon number $b_i$ in the definition of the moment 
\begin{eqnarray}
J^{B...B}_{k l} &=& \frac{1}{(2l+1)!!} \sum_i \int \frac{(b_i ... b_i) d^3p}{(2\pi)^3 E_i} \nonumber \\
&\times& [m_i^2 - (p\cdot u)^2]^{l} (p\cdot u)^{k-2l} f_0^i(1\pm f_0^i) .
\end{eqnarray}
Here it is defined for $k > 2l$ and the summation is over all the components in the system.

\bibliography{basename of .bib file}

\begin{thebibliography}{99}

\bibitem{Adcox:2004mh} 
  K.~Adcox {\it et al.}  [PHENIX Collaboration],
  Nucl.\ Phys.\ A {\bf 757}, 184 (2005).
\bibitem{Adams:2005dq} 
  J.~Adams {\it et al.}  [STAR Collaboration],
  Nucl.\ Phys.\ A {\bf 757}, 102 (2005).
\bibitem{Back:2004je} 
  B.~B.~Back {\it et al.} [PHOBOS Collaboration],
  Nucl.\ Phys.\ A {\bf 757}, 28 (2005).
\bibitem{Arsene:2004fa} 
  I.~Arsene {\it et al.}  [BRAHMS Collaboration],
  Nucl.\ Phys.\ A {\bf 757}, 1 (2005).

\bibitem{Yagi:2005yb}
  K.~Yagi, T.~Hatsuda and Y.~Miake,
  Camb.\ Monogr.\ Part.\ Phys.\ Nucl.\ Phys.\ Cosmol.\  {\bf 23}, 1 (2005).
    
\bibitem{Aamodt:2010pa} 
  K.~Aamodt {\it et al.}  [The ALICE Collaboration],
  Phys.\ Rev.\ Lett.\  {\bf 105}, 252302 (2010).
\bibitem{ATLAS:2011ah} 
  G.~Aad {\it et al.}  [ATLAS Collaboration],
  Phys.\ Lett.\ B {\bf 707}, 330 (2012).
\bibitem{Chatrchyan:2012wg} 
  S.~Chatrchyan {\it et al.}  [CMS Collaboration],
  Eur.\ Phys.\ J.\ C {\bf 72}, 2012 (2012).
  
\bibitem{Ollitrault:1992bk} 
  J.~-Y.~Ollitrault,
  Phys.\ Rev.\ D {\bf 46}, 229 (1992).
  
\bibitem{Poskanzer:1998yz} 
  A.~M.~Poskanzer and S.~A.~Voloshin,
  Phys.\ Rev.\ C {\bf 58}, 1671 (1998).
  
\bibitem{MulRug98b} 
 I.~M\"uller and T.~Ruggeri,
 \textit{Rational Extended Thermodynamics} (Springer, New York, 1998).
  
\bibitem{Wang:2016opj} 
  X.~N.~Wang (ed.),
  \textit{Quark-Gluon Plasma 5} (World Scientific, Singapore, 2015).

\bibitem{Romatschke:2007mq} 
  P.~Romatschke and U.~Romatschke,
  Phys.\ Rev.\ Lett.\  {\bf 99}, 172301 (2007).
  
\bibitem{Luzum:2008cw} 
  M.~Luzum and P.~Romatschke,
  Phys.\ Rev.\ C {\bf 78}, 034915 (2008)
  Erratum: [Phys.\ Rev.\ C {\bf 79}, 039903 (2009)].
  
\bibitem{Chaudhuri:2007zm} 
  A.~K.~Chaudhuri,
  arXiv:0704.0134 [nucl-th].
  
\bibitem{Song:2007fn} 
  H.~Song and U.~W.~Heinz,
  Phys.\ Lett.\ B {\bf 658}, 279 (2008).
  
\bibitem{Denicol:2009am} 
  G.~S.~Denicol, T.~Kodama, T.~Koide and P.~Mota,
  Phys.\ Rev.\ C {\bf 80}, 064901 (2009).
  
\bibitem{Song:2009rh} 
  H.~Song and U.~W.~Heinz,
  Phys.\ Rev.\ C {\bf 81}, 024905 (2010).
  
\bibitem{Bozek:2009dw} 
  P.~Bozek,
  Phys.\ Rev.\ C {\bf 81}, 034909 (2010).
 
\bibitem{Kharzeev:2007wb} 
  D.~Kharzeev and K.~Tuchin,
  JHEP {\bf 0809}, 093 (2008).
  
\bibitem{Monnai:2009ad}
  A.~Monnai and T.~Hirano,
  Phys.\ Rev.\  C {\bf 80}, 054906 (2009).
  
\bibitem{Ryu:2015vwa} 
  S.~Ryu, J.-F.~Paquet, C.~Shen, G.~S.~Denicol, B.~Schenke, S.~Jeon and C.~Gale,
  Phys.\ Rev.\ Lett.\  {\bf 115}, 132301 (2015).
  
\bibitem{Monnai:2012jc} 
  A.~Monnai,
  Phys.\ Rev.\ C {\bf 86}, 014908 (2012).
  
\bibitem{Shen:2017ruz} 
  C.~Shen, G.~Denicol, C.~Gale, S.~Jeon, A.~Monnai and B.~Schenke,
  Nucl.\ Phys.\ A {\bf 967}, 796 (2017).
  
\bibitem{Denicol:2018wdp} 
  G.~S.~Denicol, C.~Gale, S.~Jeon, A.~Monnai, B.~Schenke and C.~Shen,
  arXiv:1804.10557 [nucl-th].
  
\bibitem{Israel:1979wp}
  W.~Israel and J.~M.~Stewart,
  Annals Phys.\  {\bf 118}, 341 (1979).

\bibitem{Monnai:2010qp}
  A.~Monnai and T.~Hirano,
  Nucl.\ Phys.\  A {\bf 847}, 283 (2010).
  
\bibitem{Tsumura:2006hn} 
  K.~Tsumura, T.~Kunihiro and K.~Ohnishi,
  Phys.\ Lett.\ B {\bf 646}, 134 (2007)
  Erratum: [Phys.\ Lett.\ B {\bf 656}, 274 (2007)].
  
\bibitem{Tsumura:2009vm} 
  K.~Tsumura and T.~Kunihiro,
  Phys.\ Lett.\ B {\bf 690}, 255 (2010).
  
\bibitem{Tsumura:2012ss} 
  K.~Tsumura and T.~Kunihiro,
  Phys.\ Rev.\ E {\bf 87}, 053008 (2013).
  
\bibitem{Eckart:1940te} 
  C.~Eckart,
  Phys.\ Rev.\  {\bf 58}, 919 (1940).
  
\bibitem{Osada:2009cc} 
  T.~Osada,
  Phys.\ Rev.\ C {\bf 81}, 024907 (2010).
  
\bibitem{Osada:2011gx} 
  T.~Osada,
  Phys.\ Rev.\ C {\bf 85}, 014906 (2012).
 
\bibitem{Osada:2014bza} 
  T.~Osada,
  arXiv:1409.6846 [nucl-th].
  
\bibitem{Bemfica:2017wps} 
  F.~S.~Bemfica, M.~M.~Disconzi and J.~Noronha,
  arXiv:1708.06255 [gr-qc].
  
\bibitem{Hiscock:1983zz} 
  W.~A.~Hiscock and L.~Lindblom,
  Annals Phys.\  {\bf 151}, 466 (1983).
  
\bibitem{Hiscock:1985zz} 
  W.~A.~Hiscock and L.~Lindblom,
  Phys.\ Rev.\ D {\bf 31}, 725 (1985).
  
\bibitem{Israel:1976tn} 
  W.~Israel,
  Annals Phys.\  {\bf 100}, 310 (1976).
  
\bibitem{Kostadt:2000ty} 
  P.~Kostadt and M.~Liu,
  Phys.\ Rev.\ D {\bf 62}, 023003 (2000).

\bibitem{Geroch:2001xs} 
  R.~P.~Geroch,
  gr-qc/0103112.

\bibitem{Van:2007pw} 
  P.~Van and T.~S.~Biro,
  Eur.\ Phys.\ J.\ ST {\bf 155}, 201 (2008).

\bibitem{Van:2011zz} 
  P.~Van,
  EPJ Web Conf.\  {\bf 13}, 07004 (2011).
  
\bibitem{Van:2011yn} 
  P.~Van and T.~S.~Biro,
  Phys.\ Lett.\ B {\bf 709}, 106 (2012).
  
\bibitem{Becattini:2014yxa} 
  F.~Becattini, L.~Bucciantini, E.~Grossi and L.~Tinti,
  Eur.\ Phys.\ J.\ C {\bf 75}, 191 (2015).
  
  \bibitem{Onsager1}
  L.~Onsager,
  Phys.\ Rev.\ {\bf 37} 405, (1931).
  \bibitem{Onsager2}
  L.~Onsager,
  Phys.\ Rev.\ 
  {\bf 38} 2265, (1931).

\bibitem{Hosoya:1983xm} 
  A.~Hosoya and K.~Kajantie,
  Nucl.\ Phys.\ B {\bf 250}, 666 (1985).
  
\bibitem{Buchel:2007mf}
  A.~Buchel,
  Phys.\ Lett.\  B {\bf 663}, 286 (2008).
  
\bibitem{Kovtun:2004de} 
  P.~Kovtun, D.~T.~Son and A.~O.~Starinets,
  Phys.\ Rev.\ Lett.\  {\bf 94}, 111601 (2005).

\bibitem{Monnai:2014kqa} 
  A.~Monnai,
  Phys.\ Rev.\ C {\bf 90}, 021901 (2014).

\bibitem{Bazavov:2014pvz} 
  A.~Bazavov {\it et al.} [HotQCD Collaboration],
  Phys.\ Rev.\ D {\bf 90}, 094503 (2014).
  
\bibitem{Bazavov:2012jq} 
  A.~Bazavov {\it et al.} [HotQCD Collaboration],
  Phys.\ Rev.\ D {\bf 86}, 034509 (2012).
  
\bibitem{Ding:2015fca} 
  H.-T.~Ding, S.~Mukherjee, H.~Ohno, P.~Petreczky and H.-P.~Schadler,
  Phys.\ Rev.\ D {\bf 92}, no. 7, 074043 (2015).

\bibitem{Adler:2003cb} 
  S.~S.~Adler {\it et al.}  [PHENIX Collaboration],
  Phys.\ Rev.\ C {\bf 69}, 034909 (2004);
  
\bibitem{Gunther:2016vcp} 
  J.~Gunther, R.~Bellwied, S.~Borsanyi, Z.~Fodor, S.~D.~Katz, A.~Pasztor and C.~Ratti,
  EPJ Web Conf.\  {\bf 137}, 07008 (2017)

\bibitem{Sollfrank:1990qz} 
  J.~Sollfrank, P.~Koch and U.~W.~Heinz,
  Phys.\ Lett.\ B {\bf 252}, 256 (1990).
  
\bibitem{Natsuume:2007ty} 
  M.~Natsuume and T.~Okamura,
  Phys.\ Rev.\ D {\bf 77}, 066014 (2008)
  [Erratum-ibid.\ D {\bf 78}, 089902 (2008)].
  
\bibitem{Cooper:1974mv}
  F.~Cooper and G.~Frye,
  Phys.\ Rev.\  D {\bf 10}, 186 (1974).
  
\bibitem{Teaney:2003kp} 
  D.~Teaney,
  Phys.\ Rev.\ C {\bf 68}, 034913 (2003).

\bibitem{Hirano:2005wx} 
  T.~Hirano and M.~Gyulassy,
  Nucl.\ Phys.\ A {\bf 769}, 71 (2006).
  
\bibitem{Dusling:2011fd} 
  K.~Dusling and T.~Sch\"{a}fer,
  Phys.\ Rev.\ C {\bf 85}, 044909 (2012).
  
\bibitem{Noronha-Hostler:2013gga} 
  J.~Noronha-Hostler, G.~S.~Denicol, J.~Noronha, R.~P.~G.~Andrade and F.~Grassi,
  Phys.\ Rev.\ C {\bf 88}, no. 4, 044916 (2013).
  
\bibitem{Molnar:2014fva} 
  D.~Molnar and Z.~Wolff,
  Phys.\ Rev.\ C {\bf 95}, no. 2, 024903 (2017).
  
\bibitem{Denicol:2008rj} 
  G.~S.~Denicol, T.~Kodama, T.~Koide and P.~Mota,
  Phys.\ Rev.\ C {\bf 78}, 034901 (2008).
  
\bibitem{Pu:2009fj} 
  S.~Pu, T.~Koide and D.~H.~Rischke,
  Phys.\ Rev.\ D {\bf 81}, 114039 (2010).
  
\end{thebibliography}

\end{document}